# Generalized four-point characterization method for resistive and capacitive contacts

Brian S. Kim[1,a),b)], Wang Zhou[1,a)], Yash D. Shah[1,c)], Chuanle Zhou[1], N. Işık[2,d)], and M. Grayson[1,e)]

[1]*Department of Electrical Engineering and Computer Science, Northwestern University, Evanston, Illinois 60208, USA*

[2]*Walter Schottky Institut, Technische Universität München, Garching 85748, Germany*

In this paper, a four-point characterization method is developed for resistive samples connected to either resistive or capacitive contacts. Provided the circuit equivalent of the complete measurement system is known including coaxial cable and connector capacitances as well as source output and amplifier input impedances, a frequency range and capacitive scaling factor can be determined, whereby four-point characterization can be performed. The technique is demonstrated with a discrete element test sample over a wide frequency range using lock-in measurement techniques from 1 Hz - 100 kHz. The data fit well with a circuit simulation of the entire measurement system. A high impedance preamplifier input stage gives best results, since lock-in input impedances may differ from manufacturer specifications. The analysis presented here establishes the utility of capacitive contacts for four-point characterizations at low frequency.

---

[a)] These authors contributed equally to the scientific content of this paper.
[b)] Present address: Department of Electrical Engineering, Stanford University, Stanford, California 94305, USA.
[c)] Present address: Department of Physics, University of Cambridge, Cambridge CB3 0HE, United Kingdom.
[d)] Present address: Department of Computer Engineering, Tunceli University, Tunceli 62000, Turkey.
[e)] Author to whom correspondence should be addressed. Electronic mail: mgrayson@eecs.northwestern.edu.

**I. INTRODUCTION**

Ohmic contacts are widely used in four-point electrical measurements to characterize carrier mobilities and densities in novel materials and novel electronic structures. However, ohmic contacts[1,2] can be problematic in low-density systems, and require different alloy recipes for n- and p-type systems, while for new materials the recipes may not yet be developed. Capacitive contacts, studied at low frequencies around 100 Hz[3,4] and at radio frequencies[5], on the other hand can avoid such problems, and the recipe has the advantage of being independent of both the material and carrier type. In this work, we focus on the lower frequency range which is relevant for lock-in measurement techniques. Both previous works[3,4] were unable to accurately model the entire frequency range, especially when capacitive contacts were used for both current and voltage contacts. Thus further study is required to assess the utility of capacitive contacts with an accurate model that can fit in the entire frequency range, guiding the eventual design of capacitive contact samples.

In this paper, both ohmic contacts and capacitive contacts are studied and compared to simulations using the PSpice circuit simulation software. Circuit models of each component of the measurement system are first determined, including input and output impedances of the electronic measurement instruments, BNC adapters, and coaxial cables. Because the lock-in input impedance deviates significantly from equipment specifications, use of a high impedance input preamplifier is preferred to simplify circuit analysis. A discrete element test sample is then used to demonstrate this four-point characterization method for generalized contact impedances, comparing with the PSpice circuit simulation assembled from the circuit models of each component. The PSpice model well represents the behavior of the measurement system, even when capacitors were used for all four contacts. A measurement frequency range is defined within which the frequency response gives zero phase distortion, and a capacitive scaling factor is calibrated for each measurement configuration.

The remainder of this paper is organized as follows. Section 2 discusses the specific process of modeling each component in the lock-in measurement system. Section 3 then introduces a discrete element test sample used to demonstrate the generalized four-point characterization method. The experimental results of this test sample and PSpice circuit simulations of the lumped-element-modeled lock-in measurement system are compared in Section 4, as well as the definition of the measurement frequency range and scaling factor. The paper ends with concluding remarks in Section 5, and the Appendix illustrates the importance of including the high input impedance preamplifier stage.

## II. CIRCUIT EQUIVALENT OF THE LOCK-IN MEASUREMENT SYSTEM

### A. Circuit equivalent modeling procedure

To model each lock-in measurement system below, it is decomposed into its respective components, and a candidate circuit equivalent is assigned to each target component. Optimal model parameters are chosen such that the PSpice circuit simulation of the measurement system accurately models the experimental results over the range 1 Hz – 10 kHz. If the target component has more than one element in its circuit equivalent, multiple measurements and simulations are compared to deduce optimal model parameters. If the simulation cannot model the required frequency range, a new circuit equivalent is tested and the process repeated until the appropriate circuit equivalent is found. We note that we were able to get reasonable results in the experiments below only when the entire experiment was shielded, so all discrete elements were put inside aluminum boxes with BNC feedthroughs.

*1. Lock-in input*

First, we will model the lock-in input impedance itself using the simplified lock-in measurement system shown in Fig. 1(a) with circuit equivalent in Fig. 1(b). The measurement system consists of a lock-in voltage source $V_S = 1$ V; a 1 meter coaxial cable with capacitance $C_{1m} = 98$ pF; a source output impedance metal film resistor $R_S = 100$ MΩ with residual parallel capacitance $C_S = 0.17$ pF shielded inside an aluminum box; a Male-Male (MM) adapter with capacitance $C_{MM} = 2$ pF; and the lock-in input. The exact values listed above were arrived at through an iterative process to be described shortly in Section 2.2. We first model the Stanford Research 830 (SR830) lock-in input impedance according to the nominal circuit equivalent specified by the manufacturer, as shown in Fig. 2(a) with an input impedance of $R_{in} = 10$ MΩ in parallel with an input capacitance of $C_{in} = 25$ pF.

Fig.2(b) shows that PSpice simulations based on the nominal lock-in model have the correct low-pass filter behavior, but with a cut-off frequency that is higher than that observed in the experimental data. The result is highly unsatisfactory, with a factor of 3 error in the estimated $V_A$ magnitude at high frequencies. Thus we conclude that the manufacturer specification for the input impedance is not sufficiently exact for the precision analysis we wish to perform.

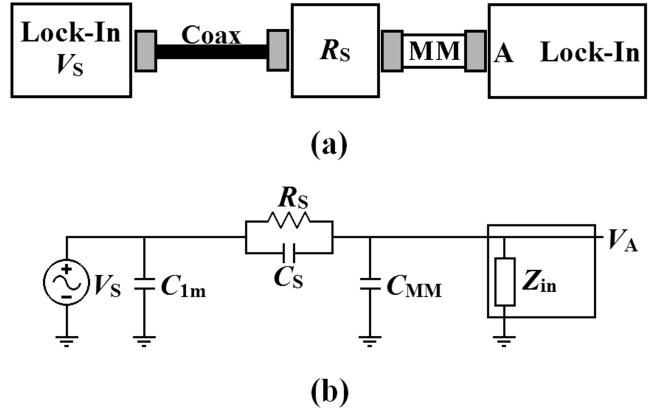

(a)

(b)

FIG. 1. (a) Diagram of the lock-in measurement system used to test the nominal lock-in input model. (b) Circuit equivalent of the same measurement system. $C_{1m}$ represents a capacitance of 1 m coaxial-cable; $R_S$ and $C_S$ the resistance and parallel capacitance of a source impedance resistor inside an aluminum shielded box; and $C_{MM}$ the capacitance of a MM connector. The lock-in input impedance $Z_{in}$ is considered for two different cases in Figs. 2 and 3.

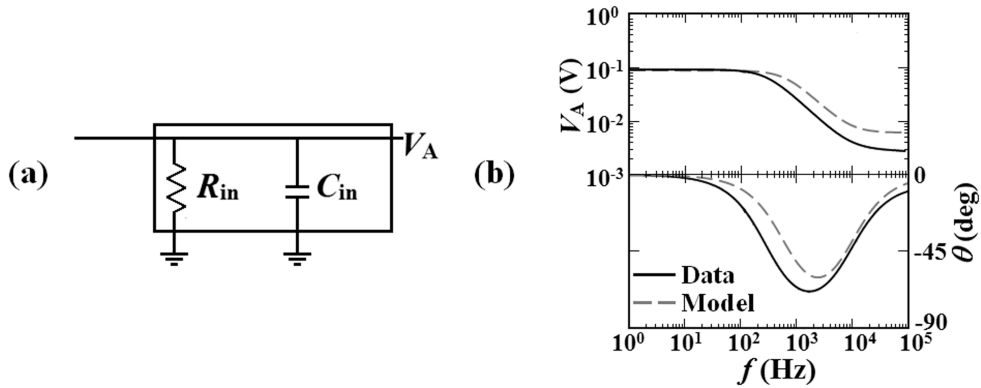

FIG. 2. (a) Nominal lock-in input impedance model. (b) Magnitude and phase plot of the lock-in input voltage $V_A$ and its corresponding PSpice circuit simulation based on the nominal lock-in input model.

But if we assume an additional low-pass current divider stage with a $R_{div}$ = 6.5 kΩ series resistor and $C_{div}$ = 31 pF parallel capacitor at the front end of nominal lock-in input stage as in Fig. 3(a), the model accurately matches the observed behavior up to 100 kHz as seen in the Bode plot of Fig. 3(b). The simulation improves since additional series resistor forms an RC circuit with $C_{in}$ and provides additional negative phase shift at high frequencies above 10 kHz. And additional parallel capacitor accounts for steeper roll-off. Therefore we will model the SR830 lock-in input as a nominal lock-in input stage with this empirical low-pass current divider at its front end.

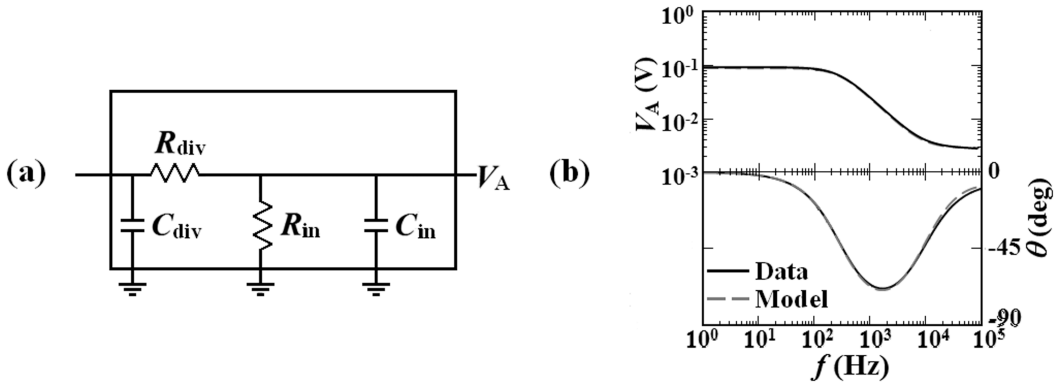

FIG. 3. (a) Modified lock-in input model with empirical low-pass current divider stage at front end. (b) Magnitude and phase plot of $V_A$ and its corresponding PSpice circuit simulations based on modified lock-in input model.

## B. Base System

The second measurement circuit of interest will allow us to model the impedance of any elements with a female or male BNC end. We call such a measurement system the *Base System,* and it can have either a male or female input, where target impedances can be inserted and measured. Fig. 4(a) shows a diagram of the Male input (M-Input) system, which is a combination of a MFF connector and an MM adapter, where one remaining F-connector serves as the input to the system. Similarly, the MMM connector functions as the female input (F-Input) system in Fig. 4(b).

Inserting a target test element to such a *Base System* input will add another impedance $Z_{test}$ to the lumped-element-modeled *Base system* in Fig. 4(c). Therefore, the impedance of the target element can be modeled by finding the resistor and capacitor values required to fit the simulation to the experimental results.

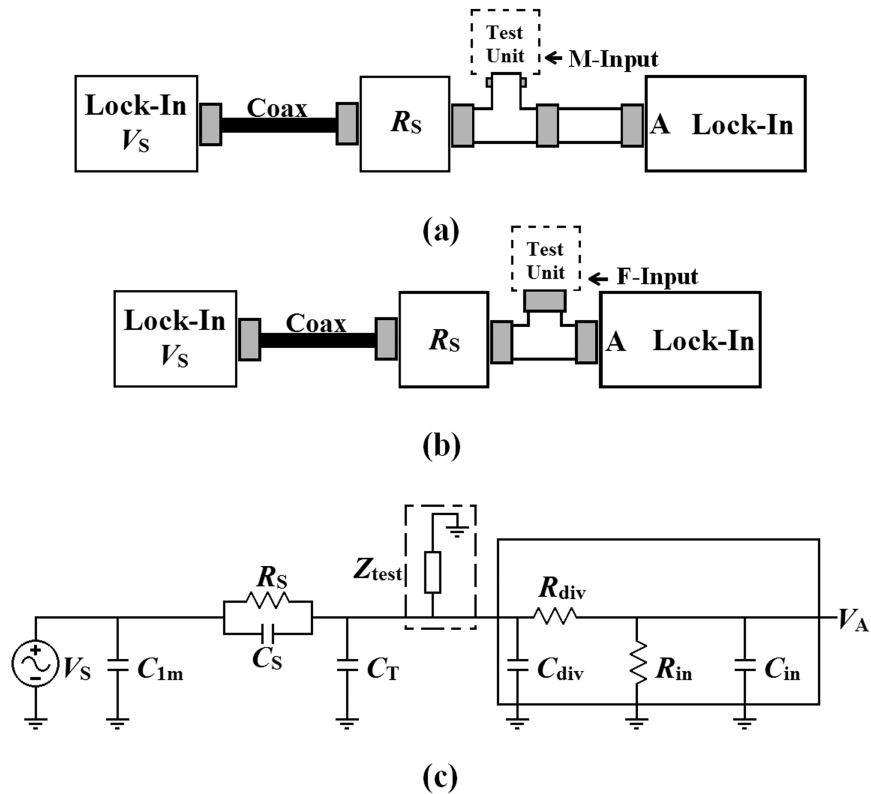

FIG. 4. (a) Diagram for testing impedance of circuit elements $Z_{test}$ with the male BNC, M-Input

*Base System*. (b) Diagram for testing $Z_{test}$ with female BNC, F-Input *Base System*. (c) Circuit equivalent of the *Base System* with a target element impedance $Z_{test}$. For the M-Input panel (a) $C_T = C_{MFF} + C_{MM} = 5$ pF, and for the F-Input panel (b) $C_T = C_{MMM} = 7$ pF.

## *1. Various capacitors*

This procedure was used to model the capacitance of various connectors in the measurement circuit in Fig. 5.

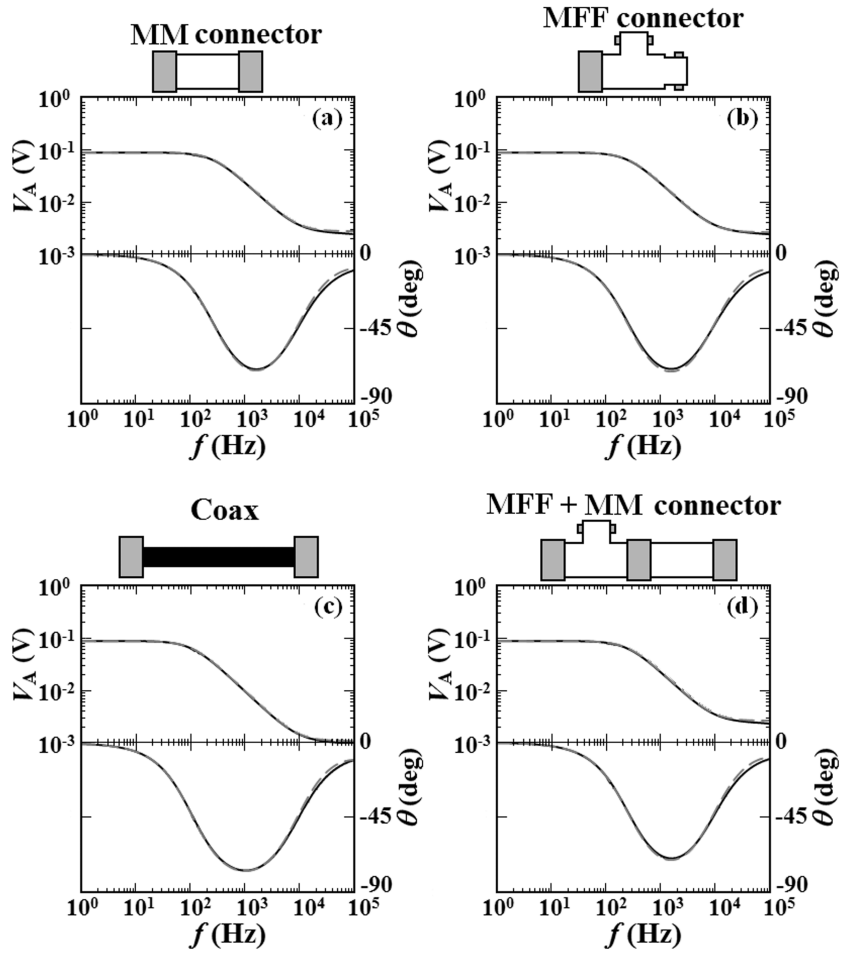

FIG. 5. Magnitude and phase plot of $V_A$ and its corresponding PSpice circuit simulations: (a) MM adapter, $C_{MM} = 2$ pF, (b) MFF connector, $C_{MFF} = 3$ pF, (c) 1m coaxial cable, $C_{1m} = 98$ pF and (d) a combination of MFF and MM connector, $C_{MFF} + C_{MM} = 5$ pF.

## III. FOUR-POINT SAMPLE WITH GENERALIZED CONTACT IMPEDANCE

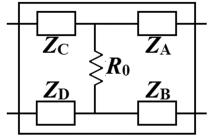

FIG. 6. Schematic of the test sample.

The discrete element four-point sample in Fig. 6 represents a generalized sample with either resistive or capacitive contact impedances, where the test resistance to be measured is labeled $R_0$. The voltage contacts with the subscript A, B and the current contacts are labeled with the subscript C, D. Therefore, four distinct configurations of four-point characterization setup are possible, as the current or voltage contacts can be either resistive or capacitive, respectively. A shorthand notation for each configuration is: Ω-Ω, Ω-κ, κ-Ω, κ-κ. The first symbol represents whether the current contacts ($Z_C$ and $Z_D$) are ohmic or capacitive, and the second symbol represents whether the voltage contacts ($Z_A$ and $Z_B$) are ohmic or capacitive, where Ω represents resistive and κ capacitive, respectively. A metal-film resistor is used for the test resistance $R_0 = 30$ kΩ. Resistive contacts are modeled with current contact resistance $R_I = R_{C,D} = 75$ kΩ and voltage contact resistance $R_V = R_{A,B} = 75$ kΩ metal film resistors, and capacitive contacts are modeled with current contact capacitance $C_I = C_{C,D} = 0.68$ nF and voltage contact capacitance $C_V$ $= C_{A,B} = 0.1$ nF polyphenylene sulphide film capacitors, respectively. These values are chosen to approximate to those of the test device published in Ref. 3.

**A. Four-point characterizations of resistive and capacitive contacts**

For the lock-in measurement system of Fig. 7(a), we included a high input impedance Ithaco 1201 preamplifier which has orders of magnitude larger input impedance than that of the lock-in, and which matches the manufacturer's nominal input specifications with pre-amp input $R_P = 100$ GΩ and $C_P = 20$ pF. By minimizing the leakage current through the voltage contacts of the test sample, the pre-amp widens the measurement frequency range for making four-point characterizations. By comparison, same measurement performed without the pre-amplifier in Appendix A is shown to be harder to interpret.

The procedure of generalized four-point characterization of $R_0$ is as follows. $V_S$ and $R_V$ in series form a current source, which sends current $I_S = V_S/R_S$ through the current contacts of the four-point sample. The measured voltage difference between two voltage contacts of the four-point sample is then divided by this source current $I_S$ flowing through $R_0$ to determine the four-point measured resistance $R_{4pt} = R_{\varrho\text{-}\varrho}, R_{\kappa\text{-}\varrho}, R_{\varrho\text{-}\kappa}, R_{\kappa\text{-}\kappa}$ which is then plotted in a Bode plot. All components of the measurement system, including resistors and capacitors of the test sample, are modeled based on methods described in Section 2.

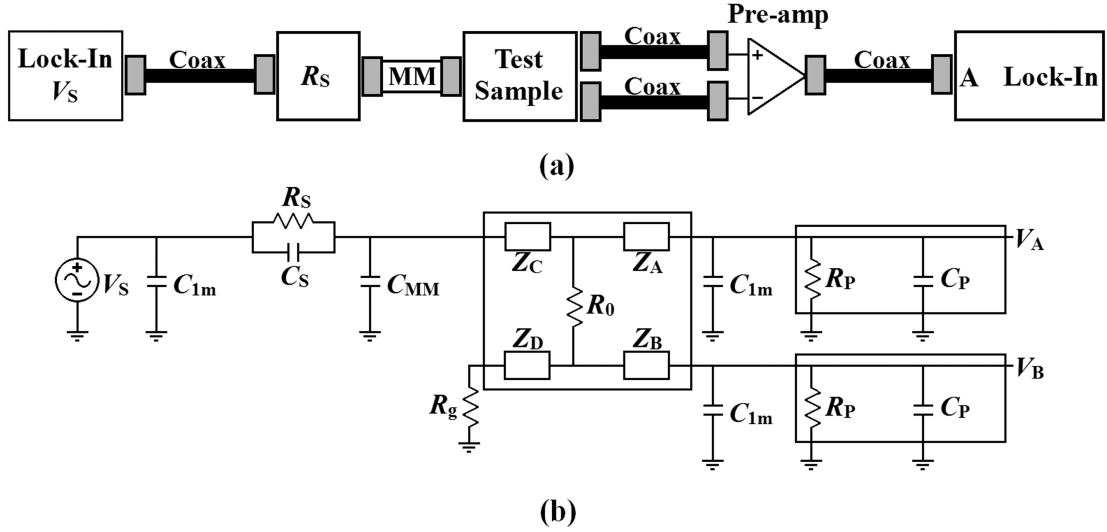

FIG. 7. (a) Diagram of the lock-in measurement system of four-point sample including preamplifier stage. (b) Circuit equivalent of the measurement system. $R_g = 50\ \Omega$ represents a resistance of ground resistor; and $R_P$ and $C_P$ represent the input resistance and capacitance of the preamplifier input.

## IV. EXPERIMENTAL RESULTS

Figure 8 shows the experimental results of the four different four-point current-voltage configurations, $\Omega$-$\Omega$, $\kappa$-$\Omega$, $\Omega$-$\kappa$, and $\kappa$-$\kappa$ in panels (a)-(d), respectively. It is clear that all four show a different frequency response, yet it is possible to deduce the test resistance $R_0$ from each of these datasets with knowledge of the measurement circuit. In the following, we first describe the frequency range wherein the $R_0$ value can be characterized, and then we specify a scaling factor which can be deduced from the measurement circuit which allows the value of $R_0$ to be determined.

## A. Measurement frequency range

The measurement frequency range is defined over the set of frequencies where the four-point voltage is directly proportional to the test resistance $R_0$. The phase of the four-point voltage is implicitly zero over this measurement range. Two particular frequencies, the low cut-off frequency $f_L$ and high cut-off frequency $f_H$, are used as the boundaries of this measurement frequency range.

As seen in Fig. 8, all the magnitude plots start to rise near 10 kHz. This increasing response is caused by the parallel source capacitance $C_S$, which starts to short the source resistance at the high cut-off frequency for all four configurations

$$f_H = (2\pi R_S C_S)^{-1} \quad (1)$$

which gives us $f_H = 9.4$ kHz, matching the results in Fig. 8.

The low cut-off frequencies for Ω–Ω and Ω–κ test samples, are 0 Hz because of the low-pass frequency response for both. The low frequency slope in the Bode magnitude plot for κ–Ω in Fig. 8(b) results from increased current due to decreasing impedance of the current capacitor $C_C$ when it is comparable with 100 MΩ = $R_S$. So the low cutoff frequency for κ–Ω test sample is 2.2 Hz, calculated as

$$f_L = (2\pi R_S C_C)^{-1} \quad (2)$$

For κ–κ mode, the circuit analysis indicates that the low cutoff frequency is

$$f_{\text{L}} = (2\pi R_{\text{S}}(C_{\text{C}} \, \text{P} C_{\text{A}} \, \text{P}(C_{1m} + C_{\text{P}})))^{-1} \qquad (3)$$

where the expression $X \| Y = XY/(X+Y)$. So the low cutoff frequency for κ–κ test sample is 32 Hz. One can see that the phase is truly zero within the measurement range for all configurations except k-k, and this phase rotation would approach zero if the $f_{\text{H}}$ high frequency cut-off were to be increased by reducing either the source resistance $R_{\text{S}}$ or capacitance $C_{\text{S}}$. The measurement frequency ranges are marked with black bars in Fig. 8, with end-points set by the low and high cut-off frequencies.

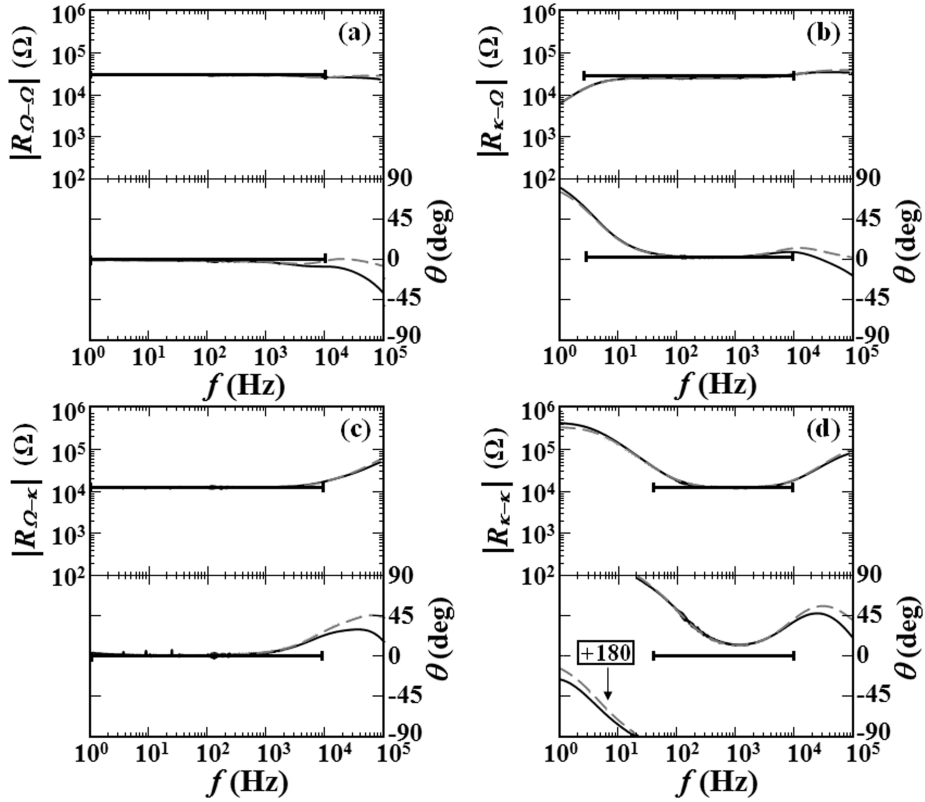

FIG. 8. Magnitude and phase Bode plots of four-point electrical impedance and its corresponding PSpice circuit simulations with lock-in preamplifier stage: (a) Ω-Ω, (b) κ-Ω, (c) Ω-κ, and (d) κ-κ.

The black bars indicate the measurement frequency range for each setting. The vertical position of the bars shows the magnitude and phase of calculated resistance, where the current divider scaling factor is included. Excellent agreement is shown for both the measurement frequency range, and for the scaling factor.

**B. Capacitive scaling factor**

When capacitive contacts are used, the network of capacitors acts as a current divider or voltage divider at various branches of the measurement circuit, causing the measured signal to be reduced by a calibrated value, called the capacitive scaling factor[6]. The capacitive scaling factor $\gamma$ is defined as the ratio of the four-point resistance $R_{4pt}$ within the measurement frequency range to the test resistance of interest $R_0$, $\gamma = R_{4pt}/R_0 < 1$. The black bars in Fig. 8 illustrate the value of $R_{4pt}$ after being multiplied by the scaling factor. Note below that although the current capacitors $C_I$ are nominally equal $C_C = C_D$, as are the voltage capacitors $C_V$ such that nominally $C_A = C_B$, eventual process variation might cause minor differences in a real sample, so the explicit notations will be used in the equations below.

(1) Ω–Ω measurement

For the Ω–Ω measurement, there are no capacitors which affect the four-point measurement, so the capacitive scaling factor is simply 1, $\gamma = 1.0$.

(2) κ–Ω measurement

For the κ–Ω measurement, $C_{MM}$ together with $C_C$ acts as a current divider, and another current division happens at input end of $R_0$, so the scaling factor calculated below gives us $\gamma \approx$

0.91:

$$C' = C_C \, P(C_D + 2(C_{1m} + C_P))  \tag{4}$$

$$\gamma = \frac{C'}{C' + C_{MM}} \times \frac{C_D + (C_{1m} + C_P)}{C_D + 2(C_{1m} + C_P)} \tag{5}$$

Here the second fraction above constitutes an additional factor added to the result derived in Ref. 3. This is important when $C_D$ is comparable the total capacitance of coaxial cable plus input capacitance of the pre-amplifier.

(3) Ω–κ measurement

For the Ω–κ measurement, there is a voltage division between $C_A$ or $C_B$ and the input capacitance of the pre-amplifier plus coaxial cable, so the capacitive scaling factor for Ω–κ is γ ≈ 0.45, calculated as

$$\gamma = \frac{C'''}{C''' + C_{MM}} \times \frac{C_D + C''}{C_D + 2C''} \times \frac{C_A}{(C_{1m} + C_P) + C_A} \tag{6}$$

(4) κ–κ measurement

For the κ–κ measurement, the current is divided between $C_{MM}$ and $C_C$, and between $C_A$ and $R_0$. The capacitive scaling factor for κ–κ measurement is γ ≈ 0.43, calculated as

$$C'' = C_A \, P(C_{1m} + C_P) \tag{7}$$

$$C''' = C_C \, P(C_D + 2C'') \tag{8}$$

$$\gamma = \frac{C'''}{C''' + C_{MM}} \times \frac{C_D + C''}{C_D + 2C''} \times \frac{C_A}{(C_{1m} + C_P) + C_A} \tag{9}$$

This formula shows similar dependence on capacitors as in the circuit of Ref. 3, however, $C_C$ and $C_D$ may not be the same because of fabrication variation, so they are referred to separately. $C_{MM}$ is included here since we used a single male-to-male adaptor which has very small but measurable capacitance, causing a small amount of current loss as a current divider. A scaling factor $\gamma$ much closer to unity results here, indicating that the experimental four-point measurement is much closer to the bare test resistance than in Ref. 3.

**C. Limitations**

The analysis of the measurement frequency range and scaling factor enlightens us on the design of resistive and capacitive contacts, in order to get a wider measurement frequency range and a scaling factor closer to 1, comparable to traditional ohmic-contact four point characterization.

(1) <u>Sample resistance.</u> The assumption has been made that the sample test resistance $R_0$ is negligible compared to the impedance of capacitors when deducing the scaling factor. A complete circuit analysis indicates that the maximum resistance for $R_0$ is

$$R_{max} = 10\% \times (2\pi f_H (C_A \, P(C_{1m} + C_P)))^{-1} \qquad (10)$$

which gives us the maximum resistance of 31.4 kΩ in our system setting. If the resistance exceeds this value, one must reevaluate the scaling factor and cut-off frequency.

(2) <u>Capacitive contact</u>. Based on our frequency response analysis, it is believed that larger current capacitors can lower $f_L$, widening the measurement frequency range. On the other hand, too large current capacitors may introduce a low-pass filter that lowers $f_H$ caused by the parallel source impedance capacitor $C_S$. If the capacitive contacts are lithographically fabricated on the sample, the percentage of sample area dedicated to the capacitive contact becomes another factor for consideration. So there is a design trade off when making capacitive contacts.

(3) <u>Capacitance of coaxial cables</u>. The capacitance of coaxial cables is a non-negligible cause of current loss due to the current divider circuit equivalence, especially when all contacts are capacitors. Therefore, shorter coaxial cables are encouraged to introduce smaller capacitances, and whenever possible, the short BNC connectors are to be used since they have only several pF capacitances. If the input capacitance of the pre-amplifier is negligible, the capacitance of the coaxial cables strongly affects the voltage distribution between voltage capacitors and inputs of pre-amplifier. If the capacitance of coaxial cables can be limited under 10% of voltage capacitors (in our system setting, this limit would be 10 pF), the voltage divider effect at the pre-amp stage can be ignored. Alternately, one can make larger voltage capacitors to allow for larger coaxial capacitances, but once again this will be limited by the sample area the final lithographic design.

## V. CONCLUSION

Results of four-point sample measurements show that the circuit equivalent of the full lock-in measurement system can accurately estimate the four-point characterization for generalized contact impedances below 100 kHz. Therefore, if capacitive scaling factors are calibrated accordingly within corresponding measurement frequency range, the use of capacitors as contacts in quantitative four-point characterizations is viable.

## APPENDIX: FOUR-POINT CHARACTERIZATION WITHOUT PREAMPLIFIER

We studied characterization of the four-point sample in a standard lock-in measurement system without using a preamplifier as shown in Fig. 9, to verify the validity of the circuit equivalent model. The comparison of experimental results and PSpice circuit simulations in Fig. 10 confirms that the circuit equivalent performs well below 100 kHz. Note that the $\kappa$–$\kappa$ configuration in Fig. 10(d) does not give a well-defined frequency range with zero-degree phase rotation. For this reason, the preamplifier stage in Section 4 is necessary to get useful results with capacitive contact measurements.

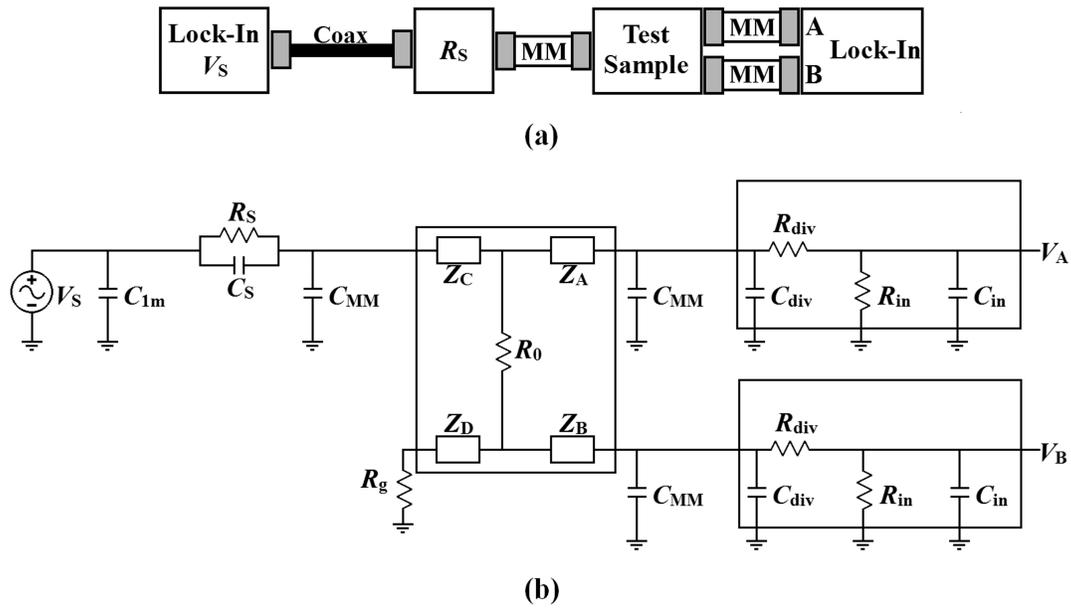

FIG. 9. (a) Diagram of the standard lock-in measurement system of test sample using the SR830 lock-in. (b) Circuit equivalent of the measurement system.

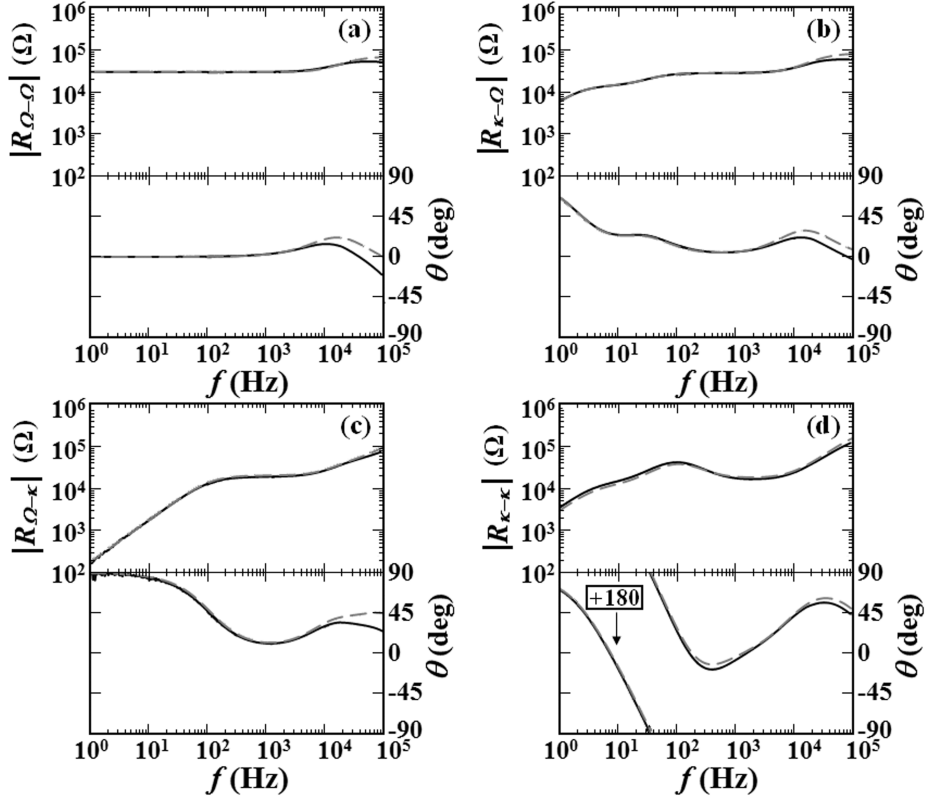

FIG. 10. Magnitude and phase plot of four-point electrical impedance and its corresponding PSpice circuit simulations for the lock-in measurement system using an SR830 lock-in: (a) Ω-Ω test sample, (b) κ-Ω test sample, (c) Ω-κ test sample, and (d) κ-κ test sample.

[1]R. H. Cox and H. Strack, Solid-State Electronics **10,** 1213 (1967).

[2]O. Göktas, Ph.D. thesis, Max Planck-Institut, Stuttgart, Germany, 2008.

[3]N. Isik, M. Bichler, S. Roth and M. Grayson, International Journal of Modern Physics B **21,** 143 (2007).

[4]Y. D. Shah, M.A. thesis, Northwestern University, Illinois, 2011.

[5]V. Dolgopolov, C. Mazur, A. Zrenner and F. Koch, J. Appl. Phys. **55,** 4280 (1984).

[6]Note that the scaling factor γ defined here is the reciprocal of that defined in Ref. 3. The present definition makes clearer the origin of the scaling factor as coming from capacitive current dividers and capacitive voltage

dividers since $\gamma < 1$.